# DESIGN OF A COMMUNICATION SYSTEM TO SEND TEXT USING LORA AT 400 MHz

Fabrizio A. Farfán[1], William C. Pérez[1], Favio D. Cabrera[1], Harold J. Carhuas and Steisy A. Carreño[1] ✉

[1] *Facultad de Ingeniería Electrónica y Eléctrica, Universidad Nacional Mayor de San Marcos, Lima, Perú*

✉ *Autor de correspondencia: Steisy Carreño Tacuri*
*Correo:* steisy.carreno@unmsm.edu.pe

*Abstract*— **This work describes the design and implementation of a low-power wireless communication system for transmitting text using ESP32 modules and the LoRa DXLR01. The proposal arises as a solution to connectivity and energy-efficiency problems commonly found in rural areas and certain urban environments where Wi-Fi or mobile networks are unavailable or operate with limitations. To address this, LoRa technology known for its long-range capability and low power consumption is integrated with an ESP32 responsible for capturing, processing, and sending messages.**

**The LoRa DXLR01 module, which operates in the 433 MHz band, is configured with parameters aimed at maximising both transmission range and efficient energy usage. Messages are sent using Chirp Spread Spectrum (CSS) modulation, improving signal penetration in obstructed areas and reducing the likelihood of errors. On the receiving end, the ESP32 interprets the data and displays it on an LCD screen. Additionally, the received information is sent to the ThingSpeak platform, allowing remote storage and visualisation without relying on conventional network infrastructure.**

**Tests conducted in a controlled environment show an average latency of 3.2 seconds for text transmission. It was also verified that the system can be used in applications such as remote monitoring, infrastructure management, and access control. However, it was observed that at distances greater than 1 km, packet loss increases, making it necessary to adjust parameters such as the Spreading Factor (SF) or transmission power for more demanding scenarios.**

## Introduction

Low-power and long-range wireless communication technologies have become essential for the development of IoT systems, especially in scenarios where traditional networks such as Wi-Fi or mobile services either perform poorly or are simply unavailable. Within this context, LoRa (Long Range) stands out as a highly efficient option for transmitting data in the sub-GHz band, as it combines low energy consumption, strong interference resistance, and a communication range that can extend several kilometres (*Find Product Documentation,* n.d.).

This project proposes the design and implementation of a point-to-point communication system to send text using the LoRa DxLR01 module, which operates in the 400–433 MHz band, together with an ESP32 and a 16×2 LCD display. The concept is simple: one node transmits messages using LoRa, and the other receives them and displays the text on the screen. Unlike Wi-Fi-based systems, no prior network or additional infrastructure is required, making it especially useful in rural, academic, or experimental environments where reliable communication is needed with limited resources.

The choice of LoRa is mainly due to its Chirp Spread Spectrum (CSS) modulation, which offers high sensitivity and strong signal penetration even at low power levels. In addition, the DxLR01 module—based on the SX1278 chip—allows configuration of parameters such as bandwidth (BW), spreading factor (SF), and coding rate (CR). This makes it possible to find an appropriate balance between range and transmission speed depending on the requirements of the environment (*Document Html | Espressif Documentation,* n.d.).

The ESP32 functions as the main control unit: it communicates with the LoRa module through SPI and also manages the display of received messages on the LCD screen. Thanks to its low power consumption and solid processing capabilities, the system is cost-effective, easy to replicate, and suitable for learning the fundamentals of long-range wireless communications. Moreover, its modular structure leaves the door open for integrating additional features such as remote monitoring, telemetry, or connection to IoT platforms.

Overall, this project serves as a practical demonstration of how LoRa and the ESP32 can be used to transmit text efficiently, securely, and with reduced energy consumption, reinforcing key knowledge in telecommunications and IoT. (*Sign Up Success - ThingSpeak IoT,* n.d.).

## BACKGROUND AND RELATED WORK

### A. Development of LoRa Communication System for Effective Transmission of Data from Underground Coal Mines

In recent years, long-range wireless communication systems have gained significant importance, mainly due to their low power consumption and their ability to transmit data even in complex environments. Among the various technologies available, LoRa (Long Range) has become a highly attractive alternative compared to options such as Wi-Fi or ZigBee, which often present limitations in range, interference, or stability (Kumar et al., 2023).

(Kumar et al., 2023) explain that because LoRa uses Chirp Spread Spectrum (CSS) modulation, it can maintain robust links while consuming very little energy, even under non-ideal conditions. This technology operates in different ISM bands such as 433 MHz, 868 MHz, and 915 MHz (Kumar et al., 2023). In their study, the authors developed a system for monitoring gases in underground mines, where they used LoRa at 433 MHz to transmit data from sensors to a surface station.

According to their results, communication remained stable up to around 20 metres under line-of-sight (LOS) conditions. In non-line-of-sight (NLOS) scenarios, the range decreased to about 13 metres before the signal was lost. Even so, the study demonstrates that LoRa can operate reliably in environments with many obstacles and high attenuation, where other technologies typically fail (Kumar et al., 2023). The system proposed by the authors followed a "star-of-stars" architecture, consisting of end nodes, gateways, and a network server. They also highlighted the importance of the 400 MHz frequency range within ISM bands, due to its better penetration in underground or high-interference environments. Based on the results obtained, the path loss exponent was approximately 1.97, indicating that propagation in confined environments shows losses even lower than those of free space due to reflection effects inside tunnels or galleries (Kumar et al., 2023)

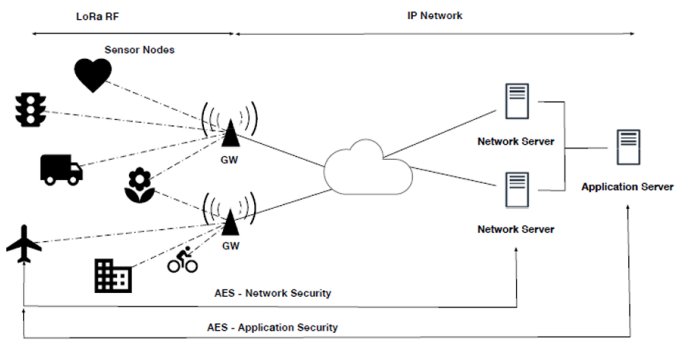

Figure 1. LoRaWAN Architecture.

These findings support the use of LoRa in applications that require long-distance data or text-message communication, with low energy consumption and without the need for complex infrastructure. The 400 MHz band, in particular, emerges as a viable option for communication projects in rural, industrial, or underground environments, given its balanced performance in terms of range, penetration, and energy efficiency (Kumar et al., 2023)

### B. A LoRa-Based Communication System for Coordinated Response in an Earthquake Aftermath

Pueyo Centelles et al. (2019) developed a communication system based on LoRaWAN aimed at improving coordination between citizens and emergency units after an earthquake. The main objective was to maintain connectivity in situations where conventional infrastructures—such as mobile networks, fibre-optic links, or copper lines—are disrupted due to structural damage or power outages. This system allows citizens to report their status through simple, predefined messages, which are then sent to authorities and rescue teams to optimise evacuation management (Centelles et al., 2019)

The proposed system was designed using a star-of-stars LoRaWAN architecture, in which end nodes communicate with multiple gateways that relay messages to the network server and the application responsible for processing them. Unlike other approaches that rely on Internet access, the system can operate autonomously thanks to its independent infrastructure and its ability to function using alternative energy sources such as batteries, which is crucial in disaster scenarios (Centelles et al., 2019)

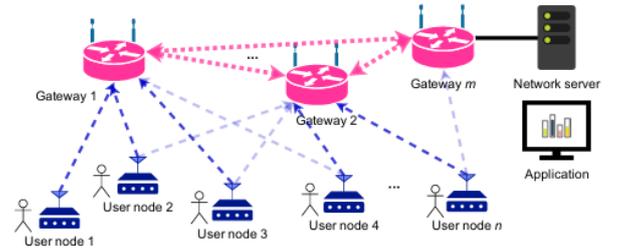

Figure 2. Architecture proposed in the article..

To validate its performance, the authors simulated a real scenario in the port city of Coquimbo, Chile, using the OMNeT++ and FLoRa tools. The model considered a population of approximately 28,000 inhabitants distributed across 7,500 households over an area of 3.5 km². The simulation results showed that the system can maintain an acceptable level of communication even when thousands of users attempt to transmit data simultaneously, although the packet delivery rate (PDR) is affected as node density increases. The study also found that performance improves significantly when optimising the number of gateways and the Spreading Factor (SF) used by end nodes (Centelles et al., 2019).

This work demonstrates the technical feasibility of LoRa as a resilient, low-power communication solution in emergency situations where other network technologies are not functional. Furthermore, its approach centred on sending short and efficient messages is applicable to projects aimed at text transmission in critical contexts, such as the one proposed in this study.

## C. A Novel LoRa LPWAN-Based Communication Architecture for Search & Rescue Missions

Manuel et al. (2022) proposed a communication architecture based on LoRa LPWAN for search and rescue (SAR) operations in scenarios where conventional infrastructures are destroyed or inaccessible. The study focused on ensuring the transmission of information between robots and base stations in environments with limited or no coverage as a result of natural disasters or large-scale accidents. For this purpose, the authors developed the Rescuer robot and the RoboMaC device, a full-duplex LoRa transceiver capable of sending control commands and receiving real-time localisation data without relying on traditional 3G, 4G, or Wi-Fi networks.

The system was designed with a device-to-device (D2D) architecture, operating within the ISM bands (902–928 MHz) and optimising parameters such as Spreading Factor (SF), bandwidth (BW), and data rate (DR) to balance range, speed, and communication reliability. The solution was validated through simulation tests (Gazebo) and real-world experiments using the Pioneer P3DX robot, achieving an effective communication range of up to 1.6 miles (approximately 2.6 km), representing a substantial improvement over Wi-Fi-based systems commonly used in emergency environments (Manuel et al., 2022).

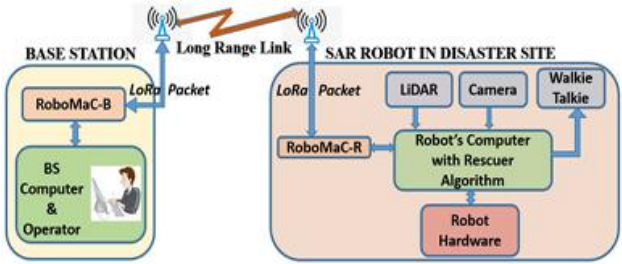

Figure 3. Block diagram of the solution.

The results demonstrated that the proposed architecture maintains a continuous link between the robot and the base station, enabling safe and coordinated operations even under complete communication blackout conditions. This work confirms the feasibility of using LoRa as a foundational technology for resilient, long-range, and low-power communication systems, making it particularly applicable to emergency management, public safety, and monitoring in critical areas such as the project developed in the present study (Manuel et al., 2022)

## D. PERFORMANCE ANALYSIS OF LORA-BACKSCATTER SYSTEMS IN AWGN CHANNELS

Ali et al. (2021) conducted a theoretical and experimental analysis of the performance of a LoRa-Backscatter (LoRa-BackCom) communication system in additive white Gaussian noise (AWGN) channels. Their proposal is framed within the search for energy-efficient solutions for small, resource-constrained Internet of Things (IoT) devices, in which battery replacement or recharging is not feasible. The research introduces an integration between LoRa technology and backscatter communication, which leverages Chirp Spread Spectrum (CSS) modulation to extend range and reduce the system's energy consumption.

The system was modelled with an uplink in which a passive tag energised by an RF source communicates with a LoRa receiver. The authors developed analytical expressions to calculate the bit error rate of the system and validated their results through numerical simulations. The study found that performance improves significantly as the Spreading Factor increases, and that the theoretical model closely matches the simulated results. Additionally, it was observed that the system's efficiency depends on the tag's reflection coefficient and the distance between the tag and the receiver, maintaining acceptable performance up to approximately 1.6 km(Ali et al., 2021).

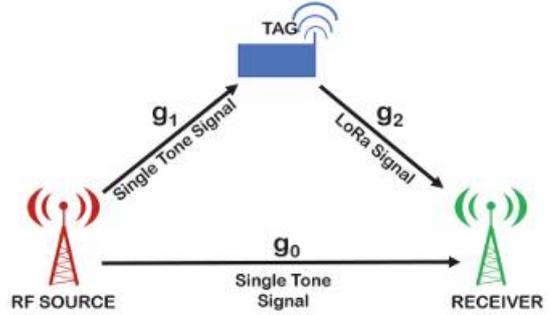

Figure 4. Enhanced BackCom system model with LoRa.

he findings of this study demonstrate that the integration of LoRa-BackCom constitutes a viable and low-power alternative for long-range communication in environments with energy constraints or without active infrastructure. This makes it a potentially applicable technology for scenarios such as environmental monitoring, public safety, and emergency response, where communication continuity is a priority (Ali et al., 2021).

## METODOLOGY

The methodology developed for this work focused on the design, implementation, and experimental validation of a point-to-point communication system based on LoRa technology operating in the 433 MHz band. The approach prioritised three fundamental aspects: low energy consumption, ease of deployment, and the ability to transmit text in hard-to-reach environments.

A. General System Architecture

The proposed system is composed of two nodes—a transmitter node and a receiver node—both based on an ESP32 microcontroller and a DxLR01 LoRa module. The link was established in peer-to-peer (P2P) mode to avoid dependency on external infrastructure.

Transmitter node: captures a predefined text message from the Arduino IDE and sends it via LoRa.

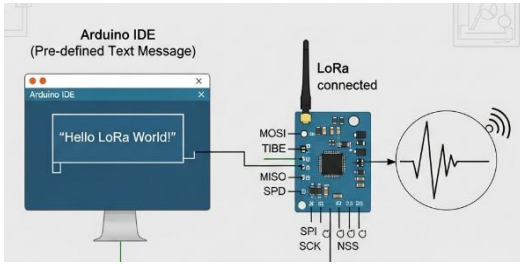

Figure 5. Reference Transmitter Node of the System to Be Implemented.

Receiver node: interprets the received frame, displays it on a 16×2 LCD screen, and subsequently sends it to the ThingSpeak platform for cloud storage.

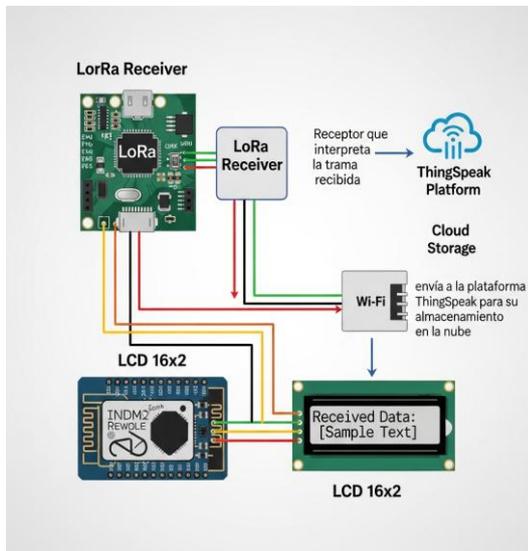

Figure 6. Reference Receiver Node of the System to Be Implemented.

This architecture makes it possible to validate the complete end-to-end flow: generation → transmission → reception → visualisation → storage.

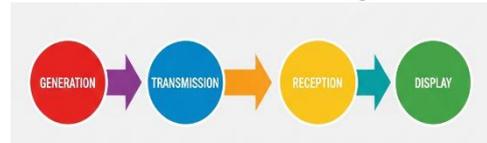

Figure 7. Flow Diagram.

B. Hardware Integration

Each node was built using low-cost, low-power components:

ESP32-DevKitC as the main controller.

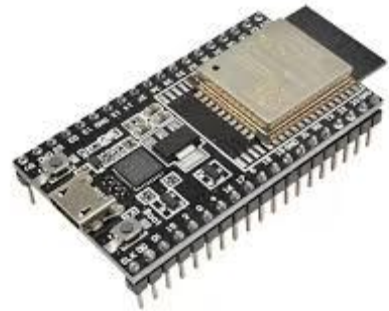

Figure 8. ESP32-DevKitC.

LoRa DxLR01 (SX1278) module, connected via the SPI interface.

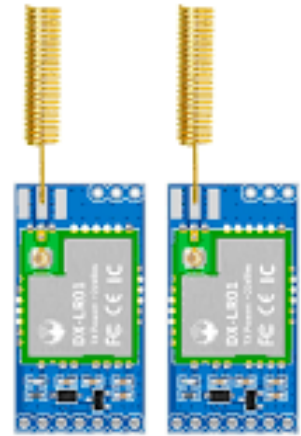

Figure 9. LoRa DxLR01 Module.

A 16×2 LCD with an I2C interface, used only in the receiver node.

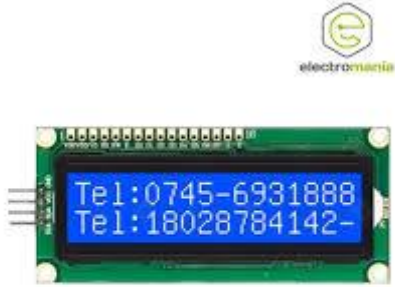

Figure 10. LCD 16×2.

The selection of the DXLR01 module is justified by its high sensitivity, ease of configuration, and compatibility with the sub-GHz band. During implementation, the correct arrangement of SPI pins (MOSI, MISO, SCK, NSS) was verified, as well as the handling of interrupts associated with the DIO0 pin, which is necessary to identify successful packet reception.

C. LoRa Parameter Configuration

To ensure a balance between range, robustness, and power consumption, the parameters shown in Table I were defined:

Table I – LoRa Configuration Parameters

| Parameter | Value |
| --- | --- |
| Frequency | 433 MHz |
| Spreading Factor (SF) | 12 |
| Bandwidth (BW) | 125 kHz |
| Coding Rate (CR) | 4/5 |
| Transmission Power | 17 dBm |

This configuration was selected based on exploratory tests. A high SF increases the receiver sensitivity and improves the achievable distance, while a moderate BW provides an acceptable compromise between data rate and robustness against interference.

D. Software and Operational Logic

The firmware was developed using the Arduino Framework, leveraging its rapid prototyping capabilities and compatibility with specific libraries for LoRa and LCD.

1. Transmitter Node

- Initialises the LoRa module with the defined parameters.
- Builds the text frame to be transmitted.
- Sends the packet using a continuous transmission function.
- Waits for a fixed delay before sending the next message.

2. Receiver Node

- Remains in listening mode ("receive mode").
- Upon detecting a valid packet, extracts the payload.
- Displays the text on the 16×2 LCD.
- Sends the message to ThingSpeak via Wi-Fi connectivity.

The use of the ESP32 enables the integration of both LoRa communication and internet connection without the need for additional hardware.

E. Ethical and Reproducibility Considerations

All parameters used, as well as the source code developed, were documented to allow full replication of the system. Restricted bands were avoided, and operation was carried out exclusively within the ISM allocations permitted for experimental communications at 433 MHz.

RESULTS

The objective of the tests was to validate whether the message transmitted from the emitter node was correctly displayed on the receiver node's LCD screen at different distances. The evaluations were conducted in a low-density urban environment, maintaining the same LoRa configuration established during the system design.

During the tests, correct message reception was verified at each point, and it was recorded whether the information appeared complete, readable, and error-free on the LCD screen. The results are summarised in Table II.

Table II – Functional Test Results

| Distance | Message Displayed on LCD |
|---|---|
| 5 m | Yes |
| 10 m | Yes |
| 20 m | Yes |
| 25 m | Yes |
| >25 m | No |

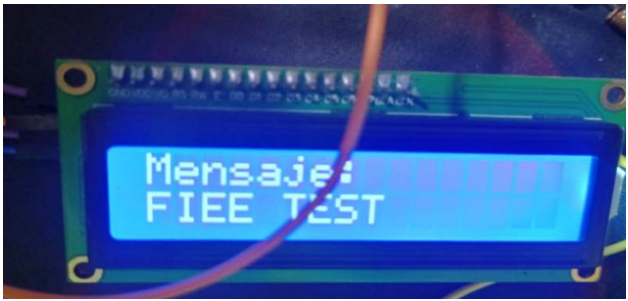

Figure 11. Test Result.

## DISCUSSION

The results obtained provide a clearer understanding of the real behaviour of the system when using LoRa as the medium for transmitting simple messages. At short and medium distances, the operation was completely stable: the message sent by the transmitter appeared on the receiver's screen without requiring retransmissions or additional adjustments. This behaviour was expected, since within this range channel attenuation is minimal and the environmental conditions do not pose a significant challenge to the signal.

As the distance between nodes increased, the influence of the environment on link performance became noticeable. Although the system continued displaying the message correctly up to approximately one kilometre, brief periods were observed where the screen did not update the received text. These interruptions did not prevent communication, but they do indicate that the link was operating near its practical limit under the test conditions.

When the nodes were positioned beyond one kilometre, communication ceased to be reliable. In several attempts, the message did not appear on the receiver, indicating that the signal was no longer strong enough to overcome the combined effects of obstacles, urban interference, and natural channel loss. This behaviour is consistent with what is commonly observed in field tests using SX1278-based modules when operating in areas with buildings or scattered vegetation.

An important point is that the system successfully met its main objective: enabling a text message to travel from the transmitter to the receiver without requiring any additional infrastructure. No gateways, repeaters, or external networks were used, demonstrating that even with a simple design it is possible to achieve functional communication over distances relevant to educational applications, basic monitoring, or scenarios where only short information transfer from one point to another is needed.

Overall, the tests confirm that LoRa technology, when properly configured, offers a viable alternative when simplicity, low power consumption, and reasonable range are required. Although it is not designed for large data volumes or highly dynamic conditions, its performance in this specific application was solid and consistent within the expected range.

## Conclusions

The work presented confirmed that LoRa technology can be effectively used to transmit simple messages in scenarios where no network infrastructure exists or where its availability is limited. The system designed based on an ESP32 and an SX1278 LoRa module successfully established a direct and stable communication link between the nodes during most of the tests, displaying the transmitted message on the receiver's LCD screen without the need for additional retransmission mechanisms.

The field tests showed that the link remains operational up to approximately one kilometre in low-density urban conditions, a sufficient distance for applications requiring basic information exchange between two separated points without relying on external services. Although performance began to degrade beyond that range, the behaviour observed is consistent with the inherent limitations of low-power devices and with the characteristics of the environment where the measurements were carried out.

Overall, the system met the objective set: demonstrating that, with an adequate configuration and accessible components, it is possible to establish functional point-to-point communication using LoRa in the 433 MHz band. This result opens the possibility for extending the work to more demanding scenarios, integrating acknowledgement mechanisms, optimising energy consumption, or evaluating performance in environments with different levels of interference. Likewise, it provides a solid foundation for educational developments and prototypes oriented towards IoT or low-bit-rate communication systems.

.